\documentclass[conference]{IEEEtran}
\usepackage{multirow}
\usepackage{float}
\floatstyle{plain}
\newfloat{bottomfigure}{bp}{lop}
\usepackage{cite}
\ifCLASSINFOpdf
  \usepackage[pdftex]{graphicx}
\else
  \usepackage[dvips]{graphicx}
\fi
\usepackage{amsmath}
\usepackage{amsthm}
\usepackage{bbm}

\usepackage{mathtools}
\usepackage{amssymb}
\usepackage[caption=true]{caption}
\usepackage[font=footnotesize]{subfig}
  \usepackage{epsfig}
\usepackage{pgfplots} 
\pgfplotsset{compat=newest} 
\pgfplotsset{plot coordinates/math parser=false} 
\newlength\figureheight 
\newlength\figurewidth 
\usepackage{color}

\newtheorem{proposition}{Proposition}

\newtheorem{theo}{Theorem}
\newtheorem{ass}{Assumption}
\newtheorem{lemma}{Lemma}
\newtheorem{remark}{Remark}

\usepackage[margin=.75in]{geometry}
\def\b0{\mbox{{\bf 0}}}
\def\bone{\mbox{{\bf 1}}}
\def\bbeta{\mbox{$\boldsymbol{\eta}$}}

\def\bgamma{\mbox{$\boldsymbol{\gamma}$}}

\def\bGamma{\mbox{$\boldsymbol{\Gamma}$}}

\def\be{\mbox{{\bf e}}}
\def\bff{\mbox{{\bf f}}}

\def\bq{\mbox{{\bf q}}}

\def\bs{\mbox{{\bf s}}}
\def\bu{\mbox{{\bf u}}}

\def\bx{\mbox{{\bf x}}}
\def\by{\mbox{{\bf y}}}

\def\bA{\mbox{{\bf A}}}

\def\bC{\mbox{{\bf C}}}

\def\bF{\mbox{{\bf F}}}
\def\bG{\mbox{{\bf G}}}
\def\bH{\mbox{{\bf H}}}
\def\bI{\mbox{{\bf I}}}
\def\bJ{\mbox{{\bf J}}}
\def\bK{\mbox{{\bf K}}}
\def\bL{\mbox{{\bf L}}}
\def\bM{\mbox{{\bf M}}}

\def\bP{\mbox{{\bf P}}}
\def\bQ{\mbox{{\bf Q}}}
\def\bR{\mbox{{\bf R}}}

\def\bbR{\mathbb{{R}}}

\def\cP{\mbox{${\mathcal{P}}$}}

\def\dbe{\mbox{$\dot{\mbox{\bf e}}$}}
\def\dbs{\mbox{$\dot{\mbox{\bf s}}$}}
\def\dbx{\mbox{$\dot{\mbox{\bf x}}$}}

\def\tgamma{\mbox{$\tilde{\bgamma}$}}

\def\hgamma{\mbox{$\hat{\bgamma}$}}
\begin{document}
\title{\vspace{18pt}Bounded Stability in Networked Systems with Parameter Mismatch and Adaptive Decentralized Estimation}
\author{Saeed Manaffam$^\dag$, Alireza Seyedi$^{\dagger\dagger}$, Azadeh Vosoughi$^\dag$, {\em Senior Member}, and Tara Javidi$^{\dagger\dagger\dagger}$, {\em Senior Member}% <-this % stops a space
\thanks{$^{\dag}$ The authors are with the Department of Electrical Engineering and Computer Sciences, University of Central Florida, Orlando, FL, emails: {\tt\small saeedmanaffam@knights.ucf.edu, azadeh@ucf.edu.}}
\thanks{$^{\dagger\dag}$ The author was with the Department of Electrical Engineering and Computer Sciences, University of Central Florida, Orlando, FL.}
\thanks{$^{\dag\dag\dag}$ The author is with the Department of Electrical and Computer Engineering, University of California at San Diego, San Diego, CA, email: {\tt\small tjavidi@ece.ucsd.edu}}
}

\maketitle
\date{\today}

%% -------------------------SECTION (I)        ABSTRACT-------------------
\begin{abstract}
	Here, we study the ultimately bounded stability of network of mismatched systems using Lyapunov direct method. The upper bound on the error of oscillators from the center of the neighborhood is derived. Then the performance of an adaptive  compensation via decentralized control is analyzed. Finally, the analytical results for a network of globally connected Lorenz oscillators are verified. 
\end{abstract}

\textbf{\textit{Index terms}:} Synchronization, complex networks, adaptive control, pinning control, parameter mismatch, parameter estimation.

%%%%%%%%%%%%%%%%%%%%%%%%%%%%%%%%
%%%%%%-----------------INTRODUCTION
%%%%%%%%%%%%%%%%%%%%%%%%%%%%%%%%
\section{Introduction}

The study of collective behavior of connected systems has drawn significant amount of attention in variety of disciplines spanning from theoretical sciences to engineering. As the study found more applications in smart grids, biological systems, etc. more scientists put effort to understand and solve the related problems in such systems \cite{Wiener65, Winfree67,Arenas08,  Watts98,Pecora98,Motter07,Manaffam13,Wang02,Zhou06,Shahrasbi11a,Shahrasbi11b,Sani14,Chen07,Manaffam13a,Yu09,Manaffam13b,Newman10}. Synchronization of these networked systems as one manifestation of such collective behavior was first introduced by Wiener \cite{Wiener65}. Pursued by Winfree in his pioneering work \cite{Winfree67}, the problem of synchronization of network of identical system was recognized as relevant in many fields of research \cite{Wiener65, Winfree67,Arenas08,  Watts98}. More recently, the introduction of the framework of master stability function by Pecora and Carroll \cite{Pecora98}, made it possible to separate the topological impact of the network from the dynamical properties of individual nodes on synchronizability of the networked systems \cite{Pecora98}\cite{Motter07}. 

Following the idea of master stability function, most of the researches have been concentrated on linking the topological properties such as minimum, maximum and average node degrees, to the synchronizability of these networks \cite{Motter07,Wang02,Manaffam13,Zhou06}. 

Since the introduction of small-worlds by Watts and Strogatz and their relations to real world networks, most of researchers turned their attentions to study the synchronizability in small world and scale-free networks \cite{Watts98,Newman10,Newman10}. Due to efficiency of dynamical flow, i.e., information flow, it has been shown that the synchronization in small-world are much easier to achieve compared to regular networks \cite{Arenas08,Manaffam13,Wu08}. In  \cite{Nishikawa03}, it has been shown that the synchronizability criteria of networks improves in homogeneous networks in contrast to heterogeneous ones.

As experimental studies have shown, similar to the network of identical systems, the network of semi-similar systems also exhibits certain collective behaviors \cite{Restrepo04, Sun09, Sorrentino11,Acharyya12,Manaffam15}. The semi-similarity in these studies implies identical structure for systems while the parameters of the systems in the network can slightly from the other systems \cite{Restrepo04, Sun09, Sorrentino11, Acharyya12,Manaffam15}. In \cite{Restrepo04}, it is reported that if in the network of semi-similar systems, if the parameters of couplings and isolated systems are slightly different, the states of all the system although cannot be absolutely equal, they can approach to a close vicinity of each other as the states evolve. The results of this work have been provided mostly on experimental merit. Following \cite{Restrepo04} and similar experimental works, a sensitivity analysis for mismatch systems and concept of $\varepsilon$-synchronization have been given in \cite{Sun09, Sorrentino11, Acharyya12}. In \cite{Sun09}, by assuming the parameter mismatch only in isolated systems, an approximate master stability function for the radius of the neighborhood which the trajectories in the network converges has been calculated. The results in \cite{Sun09} are generalized by \cite{Sorrentino11} by introducing mismatches in the inner coupling as well as weights of the connections. In \cite{Acharyya12}, a new master stability function is given by including higher terms in Taylor series of states around the average trajectories of the network. Additionally, coupling optimization to achieve ``best synchronization properties have been given \cite{Acharyya12}. The results of previous work are generalized in \cite{Manaffam15} for weighted directed systems where it has shown that the center of the neighborhood for the trajectories is the weighted average of trajectories where the weights belong to left null space of the Laplacian matrix of the network. For symmetric networks, this weighted average reduces to simple average as assumed by \cite{Restrepo04, Sun09, Sorrentino11, Acharyya12}. Also probability of $\varepsilon$-stability is used as a measure to study the phase transition of the network from desynchronization to $\varepsilon$-synchronization \cite{Manaffam15}.

In this paper, first, we investigate the problem of near synchronization in the symmetric network of mismatched systems. By using Lyapunov direct method, we calculate an upper bound on the error of trajectories from the center trajectory, where the network converge in finite time. The stated conditions on the existence of the bounded neighborhood as well as its stability are sufficient. Note that the bounds given in \cite{Sun09, Sorrentino11,Acharyya12,Manaffam15} are asymptotic bounds. Then, we choose a decentralized control to estimate and compensate for the mismatch parameters. It is shown that if each system in network has a compensator, the network will achieve absolute synchrony. Next, we choose a network of Lorenz oscillators with parameter mismatches to numerically verify our analytical results.

%%%%%%%%%%%%%%%%%%%%%%%%%%%%%%%%
%%%%%%-----------------NOTATIONS
%%%%%%%%%%%%%%%%%%%%%%%%%%%%%%%%
\subsection{Notations and Symbols}
The set of real $n$-vectors is denoted by $\bbR^{n}$ and the set of real $m\times n$ matrices is denoted by $\bbR^{m\times n}$. We refer to the set of non-negative real numbers by $\bbR_{+}$. Matrices and vectors are denoted by capital and lower-case bold letters, respectively. Identity matrix is shown by \bI. The Euclidean ($\mathcal{L}_{2}$) vector norm is represented by $\lVert\cdot\rVert $. When applied to a matrix, $\lVert\cdot\rVert $ denotes the $\mathcal{L}_{2}$ induced matrix norm, $\lVert\bA\rVert =\sqrt{\lambda_{\max}(\bA^{T}\bA)}$. Symmetric part of matrix, \bA, is denoted as 
\[\bA^{(s)}\triangleq (\bA+\bA^T)/2.\]
%%%%%%%%%%%%%%%%%%%%%%%%%%%%%%%%
%%%%%%-----------------NOTATIONS
%%%%%%%%%%%%%%%%%%%%%%%%%%%%%%%%
\section{System Model}
	Let the dynamics of networked systems is given as $ \forall i$
		\begin{align}\label{eq: NetworkEq1}
			&\dbx_i=\bff(\bx_i)+\bG(\bx_i)\delta\bgamma_i+\bu_i+\sum_{j=1}^N a_{ij}\bH(\bx_j-\bx_i)\\
			&\bx_i(t_0)=\bx_i^0. \nonumber
		\end{align}
		where $\bx_i\in\Omega$ is the state vector of node $i$, $\bff:\Omega \to \bbR^n$ describes the dynamics of the nominal system, and $\bu_i$ is the input vector. Furthermore, \bH~is inner coupling matrix and the adjacency matrix of the network is $\bA=[a_{ij}]$.  $a_{ij}\in\bbR$ is the weight of the connection from node $j$ to node $i$. There is no connection if $a_{ij}=0$. We also assume undirected connections, i.e., $a_{ij}=a_{ji}$.
		
 	Consider the nominal dynamics of each system, $i$, to be described by
	\begin{align}
			&\dbx_i=\bff(\bx_i)+\bu_i, \forall i\\\nonumber
			&\bx_i(t_0)=\bx_i^0,\nonumber
		\end{align}

	Note that $\bG(\bx_i)\delta \bgamma_i$ represents the uncertainty in the dynamics of system $i$. More precisely, $\delta\bgamma_i\in \cP$ is the uncertainty vectors of system $i$ and the uncertainty is limited to the set $\cP$ with dimension $|\cP|=m$ and uncertainty effects the individual system according to function $\bG:\bbR^n\to\bbR^{n\times m}$.  Furthermore, $\sum_j a_{ij} \bH(\bx_i-\bx_j)$ ensures that the dynamics of systems $i$ and $j$ are coupled if nodes $i$ and $j$ are connected.

Define $\bL=[l_{ij}]$ as
	\begin{align}
		l_{ij}=\left\{\begin{array}{ll}
						-a_{ij}&i\ne j,\\
						\sum_{j=1}^Na_{ij}&i=j.
					 \end{array}\right.\label{eq: Laplacian}
	\end{align}
	\bL~is a zero-sum-row matrix known as Laplacian/gradient matrix of the network and it is positive semidefinite. With this definition, \eqref{eq: NetworkEq1} can be rewritten as
	\begin{align}\label{eq: NetworkEq}
			& \dbx_i = \bff(\bx_i) + \bG(\bx_i)\delta\bgamma_i - \sum_{j=1}^N l_{ij}\bH\bx_j
			\\
			& \bx_i(t_0) = \bx_i^0. 
			\nonumber
		\end{align}
	The goal of this paper is to 1) investigate the stability properties of this network of nonlinear systems (in the absence of any input), and 2) to design and analyze the deviation from an average trajectory under a set of pinning controllers to guarantee absolute synchronization. 
	As it is known that the network mismatched system cannot be absolutely synchronized, hence, the synchronization for these systems reduces to neighborhood synchronization, where the network trajectories will converge to a certain vicinity of each other and continue to stay there\cite{Sun09, Sorrentino11, Acharyya12,Manaffam15}. To analyze this type of synchronization, the objective is to find the center and the radius of that neighborhood. In \cite{Manaffam15}, it has been shown that this center for undirected networks is simple average of all the trajectories. More precisely, define the error from the average trajectory, $\sum_{i=1}^{N}\bx_{i}/N$, as 

		\begin{align}
			\be_i\triangleq\frac1N\sum_{j=1}^N(\bx_i-\bx_j),\label{eq: Error i}
		\end{align}
		we have
		\begin{align}\label{eq: NetworkError}\dbe_i=\frac1N\sum_{j=1}^N[\bff(\bx_i)-\bff(\bx_j)]+\bG(\bx_i)\delta\bgamma_i.
		\end{align}

In the remainder of the paper, we assume the followings hold
	\begin{ass}\label{Assumption: Connected}
	The network is undirected and strongly connected.
	\end{ass}
	The connectivity of the network implies that the Laplacian matrix in \eqref{eq: Laplacian}, has only one zero eigenvalue \cite{Mohar91}. The undirectedness of the network implies the Laplacian is symmetric, hence all its eigenvalues, $\mu_{i}$, are non-negative real numbers.
	\begin{ass}\label{Assumption: BoundF}
		There exists a positive semidefinite matrix \bF~such that following inequality holds 
			\begin{align}
				(\tilde{\bx}-\tilde{\bs})^T[\bff(\tilde{\bx})-\bff(\tilde{\bs})]\le(\tilde{\bx}-\tilde{\bs})^T\bF(\tilde{\bx}-\tilde{\bs}),\label{eq: boundF}
			\end{align}
			for all $(\tilde{\bx},\tilde{\bs})\in\Omega\times\Omega$.
	\end{ass}	
	Note that this assumption is not very restrictive: If all the elements of the Jacobian of $\bff(\bx)$ with respect to state vector, \bx, is bounded, there always exists positive semidefinite matrix \bF~such that assumption \eqref{eq: boundF} holds \cite{Yu09}. This assumption is closely related to QUAD condition as discussed in \cite{DeLilles11}. Unlike QUAD condition, here \bF~is not necessarily diagonal.		

	\begin{ass}\label{Assumption: Bounded mismatch}
		There exists a symmetric positive semidefinite matrix \bGamma~and vector $\Delta\gamma$ such that following inequality holds 
			\begin{align}
				\delta\bgamma^T\bG(\bx)^T\bG(\bx)\delta\gamma\le\Delta\bgamma^T\bGamma\Delta\gamma,
			\end{align}
			for all ${(\bx,\delta\bgamma)}\in\Omega\times\cP$.
	\end{ass}	
	
Here are some Lemma which we will use in rest of the paper.
	
%	\begin{lemma}\cite[Lemma ?]{??}\label{Lemma: Joint Diagonalization}
%			If \bM~and \bK~commute, i.e., $\bM\bK=\bK\bM$, then they can be jointly diagonalized by a unitary matrix, \bQ~such that
%			\begin{align*}
%				\bM&=\bQ\bJ_M\bQ^T,\\
%				\bK&=\bQ\bJ_K\bQ^T
%			\end{align*}
%			where superscript $T$ denotes Hermitian transpose. The diagonal entries of $\bJ_M$ and $\bJ_K$ are eigenvalues of $\bM$ and \bK, respectively.
%	\end{lemma}
	
	\begin{lemma}\cite[Theorem 4. 8]{Khalil02}\label{Lemma: asymptotic W}
		 Suppose that $\bff(\bx)$ is continuous and satifies \eqref{eq: boundF} and it is uniform in $t$. Let $V:\bbR^m\to\bbR$ be contiuously differentiable function and continupous function $W(\bx)$ such that
		 \begin{align}
		 k_1\|\bx\|^{c_1}\le V(\bx)\le k_2\|\bx\|^{c_2}\\
		 \dot{V}(\bx)=\frac{\partial V}{\partial \bx}\bff(\bx)\le-W(\bx)\le 0
		 \end{align}
		 where $k_i$ and $c_i$ are positive constants. Then all solutions of 
		 \begin{align*}
			 &\dbx=\bff(\bx)\\
			 &\bx(t_0)=\bx^0,
		 \end{align*}
		 are ultimately bounded and 
		 \[\lim_{t\to\infty}W(\bx)=0.\]
	 \end{lemma}
	\begin{lemma}\cite[Theorem 4. 10]{Khalil02}\label{Lemma: Bounded x}
		 Suppose that $\bff(\bx)$ is continuous and satisfies \eqref{eq: boundF} and it is uniform in $t$. Let $V:\bbR^m\to\bbR$ be continuously differentiable function such that
		 \begin{align}\begin{array}{ll}
		 k_1\|\bx\|^{c_1}\le V(\bx)\le k_2\|\bx\|^{c_2}& \\
		 \dot{V}(\bx)=\frac{\partial V}{\partial \bx}\bff(\bx)\le-k_3\|\bx\|^{c_3} &\forall \|\bx\|\ge r\end{array}
		 \end{align}
		 where $k_i$ and $c_i$ are positive constants. Then there exists $t_1>t_0$ such that 
		 \begin{align*}\begin{array}{ll}
			 \|\bx\|\le k_4  \|\bx_0\| \exp(-c_4(t-t_0)),&\forall t_0\le t\le t_1\\
			\|\bx\|\le \left(\frac{k_2}{k_1}\right)^{1/c_1}r^{c_2/c_1}&\forall t>t_1.\end{array}
		 \end{align*}
	 \end{lemma}	
\section{Analytical Results}
In this section, first we derive the sufficient conditions on bounded stability of parameter mismatched networked systems. Then, using decentralized control, the mismatched parameters are compensated and network is driven to a reference signal, which can be a nominal or desired trajectory for the network to take.
\subsection{Boundedness Of The Synchronization Error}
	In this section, we will show that the network error of \eqref{eq: NetworkError} from the nominal invariant manifold of the network is ultimately bounded. Additionally, we will derive an upper bound on the norm of that error.
	
	\begin{theo}\label{Theorem: BoundedError}
	 Let Assumptions \ref{Assumption: Connected}-\ref{Assumption: Bounded mismatch} hold, if there exists a positive constant, $\lambda$ such that
		\begin{align}
			\bF-{\mu_i}\bH^{(s)}+\lambda\bI_{n}\prec\b0,~~~~\forall \mu_i\ne 0\label{eq: ConditionTheorem1}
		\end{align}
	 then the error from the average trajectory, \eqref{eq: Error i} is ultimately uniformly bounded as
	 \begin{align}\label{eq: Theorem1}
	 	\|\be\|\le \sqrt{ {2N}{\Delta\bgamma^T\bGamma\Delta\bgamma}/{\lambda^\star}^2}
	 \end{align}
	 where 
	 \begin{align*}
	\begin{array}{lll}
				 \lambda^{\star}=&\mbox{maximize} &\lambda  \\
				~&\mbox{subject to:} & \bF-{\mu_i}\bH^{(s)}+\lambda\bI_n\prec\b0,~~~\forall \mu_i \ne 0
				\end{array}
	\end{align*}
{\bf Proof}: see appendix \ref{Proof: BoundedError}.
	\end{theo}
		\begin{remark}
		The existence of positive $\lambda>0$ that satisfies \eqref{eq: ConditionTheorem1} is the sufficient condition on existence and convergence to the bound in \eqref{eq: Theorem1}. Additionally, according to Lemma \ref{Lemma: Bounded x}, the network reaches the bound \eqref{eq: Theorem1} in finite time.
		\end{remark}
		As oppose to previous results reported in \cite{Restrepo04, Sun09, Sorrentino11, Acharyya12,Manaffam15}, our approach guarantees convergence in finite time. Moreover, although conservative, the approach taken here does not require calculation of transition matrix of the network error to derive the conditions on the stability as well as the bound on the error, hence rendering the stability analysis of the networked systems much simpler.

\subsection{Estimation of constant uncertainty}
	In this part, we employ decentralized control to stabilize the network and compensate for the uncertainties of the network.
		
		Let \bs be the reference signal
		\begin{align}\label{eq: reference}
			&\dbs=\bff(\bs),\\
			&\bs(t_0)=\bs^0.\nonumber
		\end{align}
		
		In following theorem, we will assume that all the nodes have compensators, and the inputs are chosen as
		\begin{align}\label{eq: input}
			\bu_i&=-c_i\bH(\bx_i-\bs)-\bG(\bx_i)\hat{\bgamma}_i\\
		\label{eq: Estimation}
			\dot{\hat{\bgamma}}_i&=k_i\bG^T(\bx_i)(\bx_i-\bs).
		\end{align}
		The first term in \eqref{eq: input} is a common feedback control used in pinning control of identical networked systems\cite{Chen07,Manaffam13a,Manaffam13b,Yu09}. The second term in conjunction with \eqref{eq: Estimation} estimates and compensates for the parameter mismatches of isolated systems. The details of choosing the estimator will be given in the proof of following theorem.
		\begin{theo}\label{Theorem: MismatchEstimation}
		Let Assumptions \ref{Assumption: Connected}-\ref{Assumption: Bounded mismatch} hold. Then the network in \eqref{eq: NetworkEq} with input selected as \eqref{eq: input}, asymptotically uniformly converges to the reference signal, if the uncertainty vectors, $\bgamma_i$ are constant and there exists positive constants, $k_i>0$, and matrix $\bC=\mbox{diag}([c_1,\,\cdots,\,c_N])$, $c_i\ge0$ for all $i$ such that
				\begin{align}\label{eq: Theorem2}
					\bI_N\otimes\bF-(\bL+\bC)\otimes\bH^{(s)}\prec\b0. 
				\end{align}
		{\bf Proof}: see appendix \ref{Proof: MismatchEstimation}.
	\end{theo}
	\begin{remark}
		This theorem states that as the mismatched parameters of each isolated system is compensated by \eqref{eq: Estimation}, the problem of pinning the network to the reference trajectory \eqref{eq: reference} is reduced to the well-known pinning problem.
	\end{remark}
	There are several methods to choose the matrix \bC~to achieve synchronization. Some of these methods are given in\cite{Chen07,Manaffam13a,Yu09}.		
	 
\section{Numerical Example}
	\begin{figure}[!t]
	\includegraphics[width=8.5cm,height=7cm]{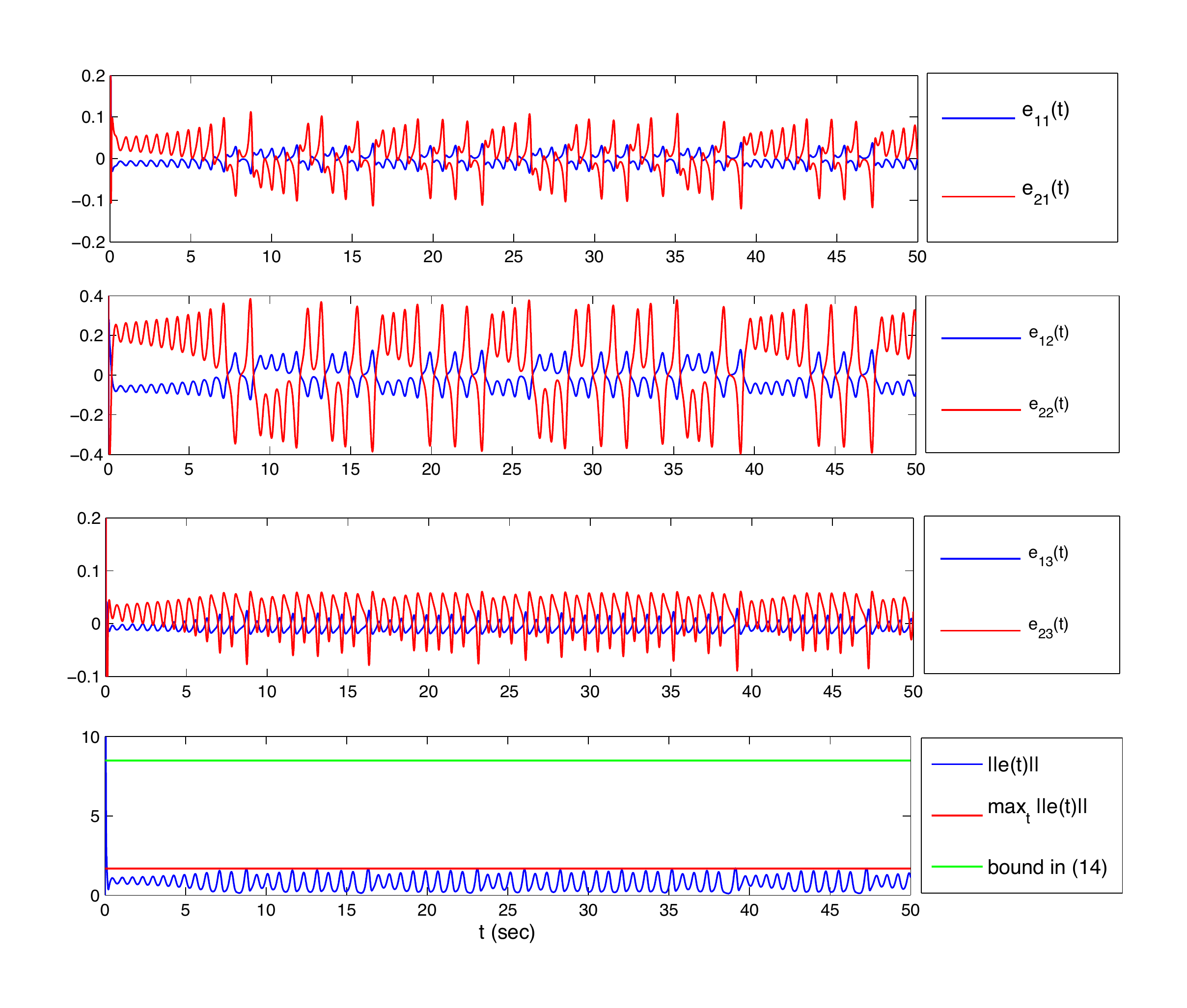}
	\caption{Norm of error for bounded uncertainties, and calculated upper bound in \eqref{eq: Theorem1}.}\label{fig: Bound}
	\end{figure}
	For the example we choose Lorenz system defined as
	\begin{align*}
	\dbx=\left[\begin{array}{c}
						a(x_2-x_1)\\
						bx_1-x_2-x_1x_3\\
						x_1x_2-cx_3
				 \end{array}\right]+\left[\begin{array}{ccc}
				 						x_2-x_1  & 0& 0\\
				 						0 & x_1 & 0\\
				 						0 & 0& -\frac83x_3
				 				 \end{array}\right]\bgamma,
	\end{align*}
	where $\bgamma=[\gamma_1~\gamma_2~\gamma_3]^T$, $(a,b,c)=(10,\,28,\,-8/3)$ and for the network we assume a globally connected network of size, $N=100$, hence $\bL=\bR_N$, $\bR_N$ is defined in \eqref{eq: Globally Connected  Net} and inner coupling matrix is assumed to be $\bH=\bI_3$. The mismatch parameters are considered to satisfy $|\gamma_{1,i}|\le 0.05a$, $|\gamma_{2,i}|\le 0.05b$, $|\gamma_{3,i}|\le 0.05c$. 

Let $\be=\bx-\bs$ then the matrix $\bF$ in Assumption \ref{Assumption: BoundF} can be computed for this case as
	\begin{align*}
		\be^T\left[\begin{array}{c}
								a(e_2-e_1)\\
								be_1-e_2-x_1x_3+s_1s_3\\
								x_1x_2-s_1s_2-ce_3
						 \end{array}\right]=&\be^T\left[\begin{array}{c}
						 								0\\
						 								-x_1x_3+s_1s_3\\
						 								x_1x_2-s_1s_2
						 						 \end{array}\right]\\&+\be^T\bM\be
	\end{align*}
	where 
	\[\bM=\left[\begin{array}{ccc}
					 -a  & a & 0\\
					 b & -1 & 0\\
					 0 & 0& -c
				\end{array}\right].\]
	\begin{align*}
		y=& e_2(s_1s_2-x_1x_2)+e_3(x_1x_2-s_1s_2)\\
		 =& e_2(s_1s_3-s_1x_3-e_1x_3)+e_3(s_1x_2+e_1x_2-s_1s_2)\\
		 =&-e_2e_3s_1-e_1e_2x_3+e_2e_3s_1+e_1e_3x_2\\
		 =&-e_1e_2x_3+e_1e_3x_2\\
	   \le& \frac{|x_3|}{2}(e_1^2/\alpha+\alpha e_2^2)+\frac{|x_2|}{2}(e_1^2/\beta+\beta e_3^2)
	\end{align*}
	Now if $|x_1|,|s_1|\le k_1$, $|x_2|,|s_2|\le k_2$, $|x_3|,|s_3|\le k_3$, we have
	\[\bF=\bM+\left[\begin{array}{ccc}
					 	\frac{k_3}{2\alpha}+\frac{k_2}{\beta}  & 0& 0\\
					 						0 & \frac{\alpha k_3}{2} & 0\\
					 						0 & 0& \frac{\beta k_2}{2}
					 \end{array}\right].\]
	With simulation we found that $k_1=20,~k_2=25,~k_3=50$, minimizing the maximum real part of eigenvalue of $\bF$ results in $\alpha=0.957$ and $\beta=3.091$. 	
	
	Substituting the values, the matrices \bF~and \bGamma, in Assumption \ref{Assumption: BoundF} and \ref{Assumption: Bounded mismatch} are 
	\begin{align*}
	\bF=&\left[\begin{array}{ccc}
					20.42  & 10.00& 0\\
					28.00 & 22.50 & 0\\
					0 & 0& 38.22
			   \end{array}\right],    \\
	\bGamma=&\left[\begin{array}{ccc}
						212.9  & 0& 0\\
						0 & 400.0 & 0\\
						0 & 0& 2500
				  \end{array}\right].
	\end{align*}

	\begin{figure}[t]
	\includegraphics[width=8.6cm,height=7cm]{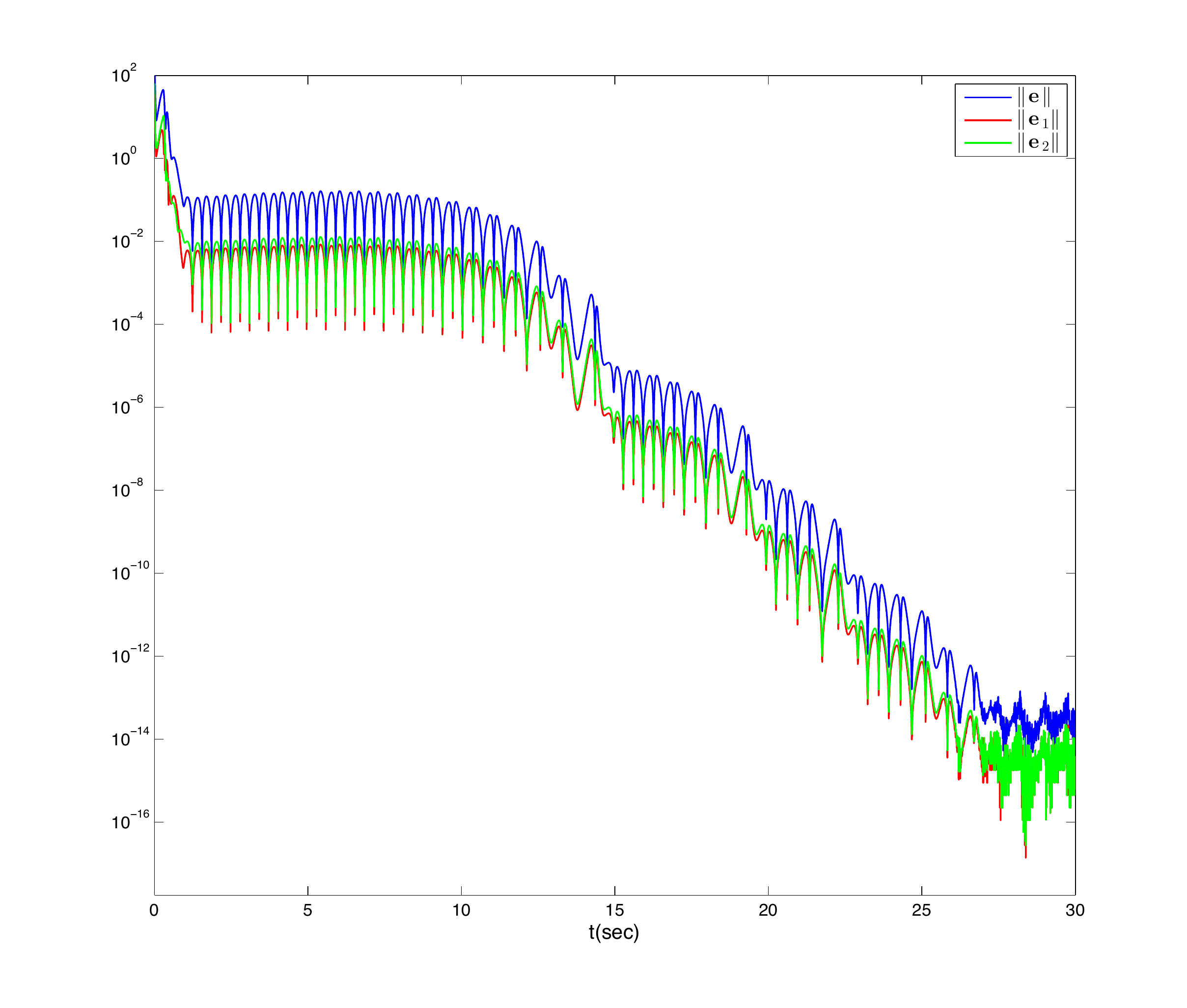}
	\caption{Norm of total error in case all the nodes are compensated/pinned, \eqref{eq: input} and \eqref{eq: Theorem2}.}\label{fig: Error_MismatchCompensated}
	\end{figure}
	
	\begin{figure}[t]
	\includegraphics[width=8.6cm,height=7cm]{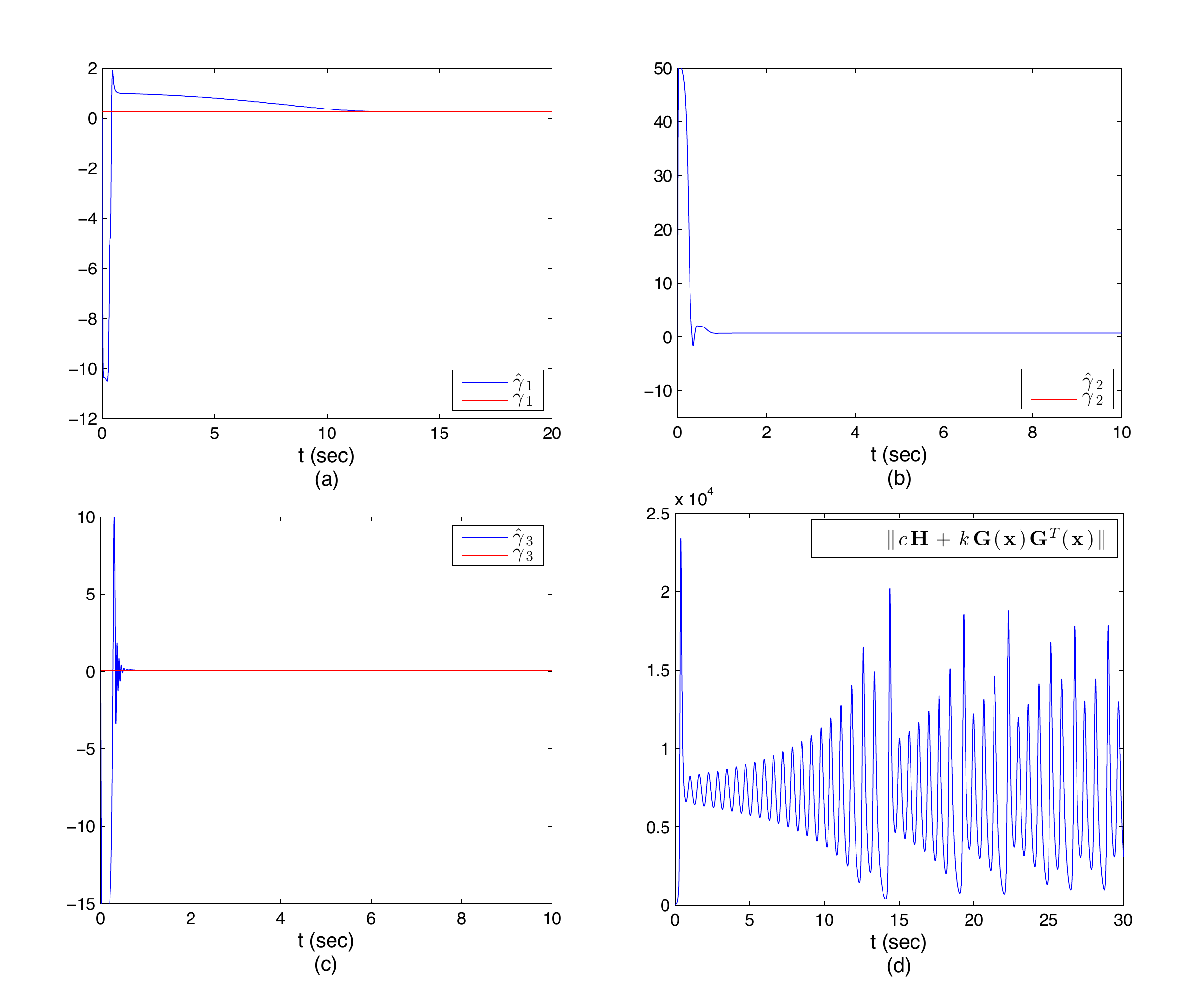}
	\caption{Estimation of uncertainties with \eqref{eq: input}.}\label{fig: MismatchEstimation}
	\end{figure}

%	\begin{figure}[t]
%	\includegraphics[width=8.8cm,height=7cm]{Fig3_theo2.pdf}
%	\caption{Sample errors from reference signal.}\label{fig: SampleErrorPinned}
%	\end{figure}

	Fig. \ref{fig: Bound} shows the simulated error of networked system from its average trajectory. As it can be seen the error from the average trajectory reaches the bound in \eqref{eq: Theorem1} in finite time as expected from Lemma \ref{Lemma: Bounded x} and Theorem \ref{Theorem: BoundedError}.

	 In the rest, we set controller's gains $c_{i}=1$ and $k_i=10$ for all $i$.\\
	The norm of sample total error for the networked system, $\|\be\|$, using \eqref{eq: input} and estimator \eqref{eq: Estimation} is given in Fig. \ref{fig: Error_MismatchCompensated}. As it can be seen, predicted by Theorem \ref{Theorem: MismatchEstimation}, the synchronization error from the reference signal damps out.
	
	Fig. \ref{fig: MismatchEstimation} gives the estimated parameters by \eqref{eq: Estimation}. The convergence of the estimated parameters can be observed from Fig. \ref{fig: MismatchEstimation}(a)-(c). The norm of overall controller gains, $\|c_{i}\bH+k_{i}\bG(\bx_{i})\bG^{T}(\bx_{i})\|$, for a sample system is plotted in Fig. \ref{fig: MismatchEstimation}(d). 

\section{Conclusion}
	Here, we have studied the ultimately bounded stability of network of mismatched systems using Lyapunov direct method. A bound on the error from the average trajectory of the networked system has been derived where the network would achieve that bound in finite time. Then, an adaptive approach control with decentralized structure has been proposed. It has been shown that the compensated network achieves absolute synchrony in presence of parameter uncertainty.

\appendices
\section{Proof of Theorem \ref{Theorem: BoundedError}}\label{Proof: BoundedError}
	\begin{IEEEproof}
	 Before proceeding to prove the theorem we need following 
	 \begin{proposition}\label{Lemma: Vector Product Inequality}
		Let \bx~and \by~to be any arbitrary vectors and \bK~to be a positive definite matrix and \bP~a matrix of proper dimensions. Then
			\begin{align*}
				\bx^T\bP\by+\by^T\bP^T\bx=2\bx^T\bP\by&\le\bx^T\bP\bK^{-1}\bP^T\bx+\by^T\bK\by.
			\end{align*}
	\end{proposition}
	\begin{lemma}\cite{Prasolov}\label{Lemma: Joint Diagonalization}
			If \bM~and \bK~commute, i.e., $\bM\bK=\bK\bM$, then they can be jointly diagonalized by a unitary matrix, \bQ~such that
			\begin{align*}
				\bM&=\bQ\bJ_M\bQ^T,\\
				\bK&=\bQ\bJ_K\bQ^T
			\end{align*}
			where superscript $T$ denotes Hermitian transpose. The diagonal entries of $\bJ_M$ and $\bJ_K$ are eigenvalues of $\bM$ and \bK, respectively.
	\end{lemma}
		Let $V(\be)=1/2\be^T\be=1/2\sum_{i=1}^N\be_i^T\be_i$, then,
		\begin{align*}
			\dot{V}=& 1/2\sum_{i=1}^N\be_i^T\dbe_i+1/2\sum_{i=1}^N\dbe_i^T\be_i									\\
				   =& \frac1{N^2}\sum_{i=1}^N\left(\sum_{j=1}^N(\bx_i-\bx_j)^T\sum_{k=1}^N[\bff(\bx_i)-\bff(\bx_k)]\right)   \nonumber   \\ 
				   	&-\sum_{i,j=1}^Nl_{ij}\be_i^T\bH^{(s)}\be_j+\sum_{i=1}^N\be_i^T\bG(\bx_i)\delta\bgamma_i 							\nonumber   
		\\
						   =&\frac1{N^2}\sum_{i,j,k=1}^N(\bx_i-\bx_j)^T[\bff(\bx_i)-\bff(\bx_k)]\nonumber  \\
				   &-\sum_{i,j=1}^Nl_{ij}\be_i^T\bH^{(s)}\be_j+\sum_{i=1}^N\be_i^T\bG(\bx_i)\delta\bgamma_i.
		\end{align*}
		Since the $\sum_{i,j=1}^{N}(\bx_{i}-\bx_{j})=\b0$, the first sum, referred to as $V_1$, can be rewritten as
		\begin{align*}
			V_1=&\frac1N\sum_{i,j=1}^N(\bx_i-\bx_j)^T\bff(\bx_i)	
		\end{align*}
		\begin{align*}
			   =&\frac1{2N}\sum_{i,j=1}^N(\bx_i-\bx_j)^T\bff(\bx_i)-\frac 1{2N} \sum_{i,j=1}^N (\bx_j-\bx_i)^T \bff(\bx_j)																  						 \\
			   =&\frac1{2N}\sum_{i,j=1}^N(\bx_i-\bx_j)^T[\bff(\bx_i)-\bff(\bx_j)]	 							\\
			   \stackrel{(a)}{\le}&\frac1{2N}\sum_{i,j=1}^N(\bx_i-\bx_j)^T\bF(\bx_i-\bx_j)										\\
			   \le&\frac1{2N}\sum_{i,j=1}^N(\be_i-\be_j)^T\bF(\be_i-\be_j)=\frac1{N}\sum_{i,j=1}^N\be_i^T\bF\be_i-\be_i^T\bF\be_j																					 \\
			   \le&\frac1N\be^T(\bR_N\otimes\bF)\be.
		\end{align*}
		where inequality $(a)$ is from Assumption \ref{Assumption: BoundF} and
		\begin{align}\label{eq: Globally Connected  Net}
				\bR_N\triangleq\left[\begin{array}{ccccc}
							N-1 &-1&-1&\cdots&-1\\
							-1 & N-1&-1&\cdots&-1\\
							\vdots&\ddots&\cdots&\vdots\\
							-1& -1&\cdots &-1 & N-1
							\end{array}\right]_{N\times N}.
		\end{align}
		Using Lemma \ref{Lemma: Vector Product Inequality} and Assumption \ref{Assumption: Bounded mismatch}, 
		\begin{align*}
			\sum_{i=1}^N\be_i^T\bG(\bx_i)\delta\bgamma_i&\le \sum_{i=1}^N\frac{\beta}2\be_i^T\be_i+ \frac1{2\beta} \delta\bgamma_i^T\bG(\bx_i)^T\bG(\bx_i)\delta\bgamma_i,			\\
			&\le\frac{\beta}2\be^T\be+\frac N{2\beta}\Delta\bgamma^T\bGamma\Delta\bgamma
		\end{align*}
		where $\beta$ is an arbitrary positive constant. 
		
		Therefore,
		\begin{align*}
			\dot{V}\le\be^T\Big(\frac1N\bR_N\otimes\bF-\bL\otimes\bH^{(s)}+\frac{\beta}2\bI_{Nn}\Big)\be
			+\frac N{2\beta}\Delta\bgamma^T\bGamma\Delta\bgamma.
		\end{align*}
		Since $\bR_N$ and \bL~ are both symmetric and Laplacian, they commute, i.e., $\bL\bR_N=\bR_N\bL$, hence by Lemma \ref{Lemma: Joint Diagonalization} there exists a unitary matrix, \bQ, such that $\bR_N$ and \bL~are jointly diagonalizable 
		\begin{align*}\begin{array}{lr}
			\bL=\bQ\bJ_L\bQ^T& \bR_N=\bQ\bJ_R\bQ^T,  \end{array}
		\end{align*}
		where $\bJ_R=\mbox{diag}([0,~N,\cdots,~N])$.
		Define
					\[\bbeta\triangleq (\bQ\otimes\bI_n)\be,\]
				then
					\begin{align*}
						\dot{V}\le&\bbeta^T\left(\frac1N\bJ_R\otimes\bF-\bJ_L\otimes{\bH^{(s)}}+\frac{\beta}{2}\bI_{Nn}\right)\bbeta\\
						          &+\frac N{2\beta}\Delta\bgamma^T\bGamma\Delta\bgamma.
					\end{align*}
						
				{As any Laplacian matrix of a connected network has one eigenvalue zero, $\bJ_R^{(N)}=\bJ_L^{(N)}=0$, with eigenvector $\bq_N=\bone_N/\sqrt{N}$, hence,}
					\[(\bq_N^T\otimes\bI_n)\be=\frac1{\sqrt{N}}\sum_{i=1}^N\be_i=\frac{\bbeta_N}{\sqrt{N}}.\]
				We know that $\sum_{i=1}^N\be_i=\b0$, thus $\bbeta_N=\b0$.
					
				Now if there exists a constant $\rho>0$ such that
						\begin{align*}
							\bF-{\mu_i}\bH^{(s)}+(\frac{\beta}{2}+\rho)\bI_{n}\prec\b0,\quad \forall i\in\{1,\,\cdots,\,N-1\}
						\end{align*}
						where $\mu_i$ are nonzero eigenvalue of the Laplacian matrix, \bL; then
						\begin{align*}
							\dot{V}(\be)\le -\rho\|\be\|^2 +\Delta.
						\end{align*}
				Using lemma \ref{Lemma: Bounded x} and setting $r^{2}=\frac N{2\beta}\Delta\bgamma^T\bGamma\Delta\bgamma/(\rho-\epsilon)$
						\begin{align*}
						\begin{array}{lc}
							\dot{V}(\be)\le -\epsilon\|\be\|^2, &\forall \|\be\|\ge\sqrt{\frac N{2\beta(\rho-\epsilon)}\Delta\bgamma^T\bGamma\Delta\bgamma},\end{array}
						\end{align*}
			Now if we set $\beta=\lambda^\star/2$, which maximizes the denominator and hence minimizes the bound on norm of error as
			\begin{align*}
				\begin{array}{lc}
					\dot{V}(\be)\le -\frac{\epsilon}{2}\|\be\|^2, &\forall \|\be\|\ge\sqrt{ {2N}{\Delta\bgamma^T\bGamma\Delta\bgamma}/{(\lambda^\star-\epsilon)}^2}.
				\end{array}
			\end{align*}
	\end{IEEEproof} 
%%%%%%%%%%%%%%%%%%%%
\section{Proof of Theorem \ref{Theorem: MismatchEstimation}}\label{Proof: MismatchEstimation}
\begin{IEEEproof}
		Let 
		\[\tgamma_i\triangleq\bgamma_i-\hgamma,\] 
		\[V=\frac12\sum_{i=1}^N\be_i^T\be_i+\sum_{i=1}^N\frac1{2k_i}\tgamma_i^T\tgamma_i.\]
		Then,
		\begin{align*}
			\dot{V}=&\frac12\sum_{i=1}^N\dbe_i^T\be_i+\be_i^T\dbe_i+\sum_{i=1}^N\frac1{k_i}\dot{\tgamma}_i^T\tgamma_i\\
				   =&\sum_{i=1}^N\be_i^T[\bff(\bx_i)-\bff(\bs)]+\be_i^T\bG(\bx_i)\tgamma_i- c_i\be_i^T\bH^{(s)}\be_i\\
				    &-\sum_{i,j=1}^Nl_{ij}\be_i^T\bH^{(s)}\be_j-\sum_{i=1}^N\frac1{k_i}\dot{\hgamma}^T_i\tgamma_i,
		\end{align*}
		substituting $\dot{\hgamma}$ from \eqref{eq: input} 
		\begin{align*}
					\dot{V}=&\sum_{i=1}^N\be_i^T[\bff(\bx_i)-\bff(\bs)]+(\be_i^T\bG(\bx_i)-\frac1{k_i}\dot{\hgamma}^T_i)\tgamma_i\\
						    &-\sum_{i,j=1}^Nl_{ij}\be_i^T\bH^{(s)}\be_j- \sum_{i=1}^Nc_i\be_i^T\bH^{(s)}\be_i\\
						 \le&\be^T(\bI_N\otimes\bF-(\bL+\bC)\otimes\bH^{(s)})\be.
				\end{align*}
		If \eqref{eq: Theorem2} holds, then using Lemma \ref{Lemma: asymptotic W}, we conclude that $\|\be\| $ uniformly goes to zero, $\|\be\|\to 0$, and $\|\tgamma_i\|$'s are bounded.
		\end{IEEEproof}
		
%%%%%%%%%%%%%%%%%%%

%

\bibliographystyle{IEEETran}
\bibliography{mybib}

\end{document}